\documentclass[aps,prb,showpacs,twocolumn]{revtex4}

\usepackage[T1]{fontenc}
\usepackage{amssymb}
\usepackage{amsbsy}
\usepackage{amsmath}             
\usepackage{epsfig}

\begin{document}

\noindent
{\bf Comment on ``Universal Spin-Flip Transition in Itinerant
  Antiferromagnets" by G. Varelogiannis:}

\vspace{0.2cm}

\begin{figure}[h]
\centerline{\psfig{file=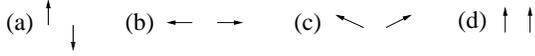,width=7cm,angle=0}}
\caption{Antiferromagnetic states: (a) parallel configuration, ${\bf
    n_r}=(-1)^{\bf r} \hat {\bf z} \parallel {\bf H}$; (b) perpendicular
    configuration, ${\bf n_r}=(-1)^{\bf r}\hat {\bf x}\perp \hat {\bf H}$; 
    (c) canted state,
    ${\bf n_r}=((-1)^{\bf r}\sin\theta,0,\cos\theta)$. (d) The normal state
    has a ferromagnetic component induced by the magnetic field (${\bf  n_r} =
    \hat {\bf z}$). } 
\label{fig1}
\end{figure}

In a recent paper,\cite{Varelogiannis03} it is argued that an
itinerant antiferromagnet in an external magnetic field undergoes a spin-flip
transition, in marked contrast with the behavior of a
localized antiferromagnet: for a weak magnetic field, the magnetization is
parallel to the field (Fig.~\ref{fig1}a), 
and flips to the perpendicular
configuration (Fig.~\ref{fig1}b) 
at a critical value of the field. A similar
spin-flip transition is predicted to occur as a function of temperature.

In this Comment we show -- considering only the zero-temperature case -- that
the conclusions of Ref.~\onlinecite{Varelogiannis03} are incorrect. The
antiferromagnetic state in the perpendicular configuration has a finite 
transverse susceptibility: a uniform magnetic field applied perpendicular to
the antiferromagnetic magnetization will inevitably induce a uniform
magnetization. As a result the energy of the canted state
(Fig.~\ref{fig1}c) will always be
lower than that of the antiferromagnetic state in the perpendicular 
configuration. The actual ground state of the system should be determined from
the free energies of the various phases that are considered (including the
normal phase). It is not sufficient, as done in
Ref.~\onlinecite{Varelogiannis03}, to find a solution
with a finite order parameter and infer the ground state from the amplitude of
the magnetization. The canted state -- not considered in
Ref.~\onlinecite{Varelogiannis03} -- turns out to be the antiferromagnetic
ground state of the system up to a critical value of the field where the normal
state is restored. 

To illustrate these points, we consider the mean-field Hamiltonian of the
two-dimensional half-filled repulsive Hubbard model in a uniform field $H$
parallel to the $z$ axis and coupled to the fermion spins:\cite{Dupuis04}  
\begin{equation}
H = -\sum_{{\bf r},{\bf r}'} c^\dagger_{\bf r} t_{{\bf r},{\bf r}'} c_{{\bf
    r}'}  
  - \sum_{\bf r} c^\dagger_{\bf r}(h \sigma^z+m \boldsymbol{\sigma} \cdot
  {\bf n_r} ) c_{\bf r} + N \frac{m^2}{U} ,
\label{ham}
\end{equation}
where $h=\mu_BH$ and $c_{\bf r}=(c_{{\bf r}\uparrow},c_{{\bf
r}\downarrow})^T$. $N$ is the total number of sites, 
$t_{{\bf r},{\bf r}'}$ a hopping integral between nearest-neighbor sites, and
$\boldsymbol{\sigma}=(\sigma^x,\sigma^y,\sigma^z)$ stands for the  Pauli
matrices. $m$ and ${\bf n_r}$ (${\bf n}^2_{\bf 
r}=1$) determine the amplitude and the direction of the magnetization,
respectively. Although written in real space, the Hamiltonian (\ref{ham}) is
similar to that considered in Ref.~\onlinecite{Varelogiannis03}. For $h=0$ and
${\bf n}_{\bf r}=(-1)^{\bf r} \hat {\bf z}$, it describes the crossover 
from a Slater ($m\sim t e^{-2\pi\sqrt{t/U}}$) to a
Mott-Heisenberg ($m\sim U/2$) antiferromagnet as $U$
increases.\cite{Borejsza04}  

We consider the three antiferromagnetic states that are schematically depicted
in Fig.~\ref{fig1}, as well as the normal state
(Fig.~\ref{fig1}d). Diagonalizing the Hamiltonian (\ref{ham}), we obtain the
free energy $F^{(\parallel,\perp)}=m^2/U-\sum_\sigma\int_{\bf k}
E^{(\parallel,\perp)+}_{{\bf k}\sigma}/2$, where $E^{(\parallel)\pm}_{{\bf
    k}\sigma} = -\sigma h \pm (\epsilon_{\bf k}^2+m^2)^{1/2}$ and
$E^{(\perp)\pm}_{{\bf k}\sigma} = \pm [(\epsilon_{\bf k}-\sigma h- \sigma m 
  \cos\theta)^2+(m\sin\theta)^2]^{1/2}$ 
are the excitation energies (obtained from the poles of the single-particle
Green function) and $\epsilon_{\bf k}=-2t(\cos k_x + \cos k_y)$ (assuming a
square lattice). The amplitude $m$ of the 
magnetization is obtained from $\partial F/\partial m=0$. $\theta$ is obtained
from $\partial F/\partial\theta=0$ in the canted state (c), whereas
$\theta=\pi/2$ in the AF state (b). In the normal phase, 
$F_N=m^2/U-\int_{\bf k}|\epsilon_{\bf k}-h-m|$. 

\begin{figure}[h]
\epsfxsize 8.cm
\epsffile[60 520 390 630]{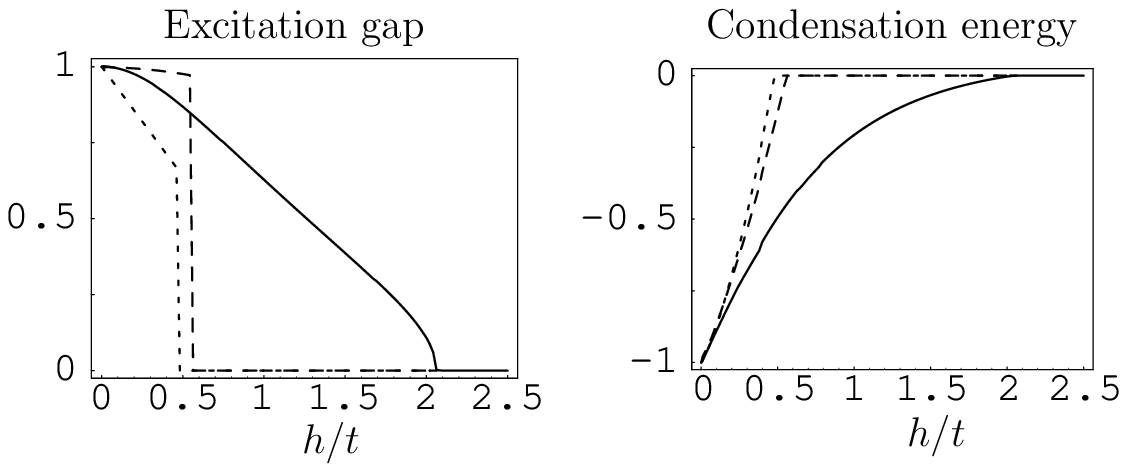}
\caption{Left panel: excitation gap in the parallel configuration (a) ($m-h$,
  short-dashed line), the perpendicular configuration (b) ($m$, long-dashed
  line), and the canted configuration (c) ($m\sin\theta$, solid line)
  [$U=4t$]. Right panel: free energy $F^{(\parallel,\perp)}-F_N$. All
  quantities are normalized to their value at $h=0$. }
\label{fig2}
\end{figure}

For the parallel (a) and
perpendicular (b) configurations, the mean-field equation $\partial F/\partial
m=0$ agrees with Ref.~\onlinecite{Varelogiannis03} and yields the same
excitation gap (Fig.~\ref{fig2}).  Although the parallel configuration has the
largest magnetization ($m$) in weak field,\cite{Varelogiannis03} it is not the
ground state. The free energies are shown in Fig.~\ref{fig2}. While the three
antiferromagnetic states (a,b,c) are degenerate when 
$h=0$, the perpendicular configuration (b) has always a lower free energy than
the parallel one (a) for any finite field, in contradiction with the
conclusions of Ref.~\onlinecite{Varelogiannis03}. Moreover, the canted state
has the lowest free energy and is therefore the actual 
ground-state. When $H$ increases, the angle $\theta$ decreases and vanishes at
the second-order phase transition to the normal phase ($h\simeq 2.06t$ in
Fig.~\ref{fig2}), in qualitative agreement with the behavior of the
magnetization in a localized antiferromagnet. 

\vspace{0.2cm}

\noindent N. Dupuis \\
Laboratoire de Physique des Solides, CNRS UMR 8502, Universit\'e
Paris-Sud, 91405 Orsay, France

\end{document}